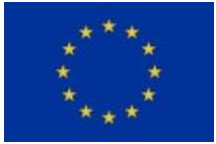
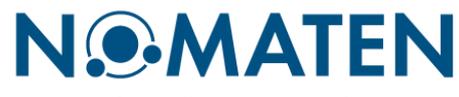

This work was carried out in whole or in part within the framework of the NOMATEN Centre of Excellence, supported from the European Union Horizon 2020 research and innovation program (Grant Agreement No. 857470) and from the European Regional Development Fund via the Foundation for Polish Science International Research Agenda PLUS program (Grant No. MAB PLUS/2018/8), and the Ministry of Science and Higher Education's initiative "Support for the Activities of Centers of Excellence Established in Poland under the Horizon 2020 Program" (agreement no. MEiN/2023/DIR/3795).

The version of record of this article, first published in Materials Characterization, Volume 203, September 2023, 113060, is available online at Publisher's website: https://doi.org/10.1016/j.matchar.2023.113060

# The effects of high-temperature ion-irradiation on early-stage grain boundaries serrations formation in Ni-based alloys

M. Frelek-Kozak[1*], K. Mulewska[1], M. Wilczopolska[1], D. Kalita[1], W. Chromiński[1,2], A. Zaborowska[1], Ł. Kurpaska[1], J. Jagielski[3,4]

1. NOMATEN Centre for Excellence - National Centre for Nuclear Research, A. Soltana 7 St., 05-400 Otwock, Poland
2. Warsaw University of Technology, Faculty of Materials Science and Engineering, st. Wołoska 141, 02-507 Warsaw, Poland
3. Łukasiewicz Research Network – Institute of Microelecotronic and Fotonics, Wólczyńska 133 St., 01-919 Warsaw, Poland
4. Material Physics Department - National Centre for Nuclear Research, A. Soltana 7 St., 05-400 Otwock, Poland

\* corresponding author: malgorzata.frelek-kozak@ncbj.gov.pl

**Abstract**

Nickel-based superalloys display outstanding properties such as excellent creep strength, remarkable fracture toughness parameters, and corrosion resistance. For this reason, Ni-based materials are considered as materials dedicated to the IV generation of nuclear reactors. Although these materials seem promising candidates, their radiation resistance and impact of radiation damage on the deformation mechanism are still not fully understood. In this work, two commercially available nickel-based alloys (Hastelloy X and Haynes 230) were investigated. Structural and mechanical properties have been described by means of SEM/EBSD, TEM, and nanoindentation tests. Radiation damage has been performed by Ar-ion with energy 320keV with two doses up to 12dpa. Obtained results have revealed a hardening effect for both levels of damage. However, more intensive effects were observed for Hastelloy X. Moreover, a significant change in precipitates' morphology in Hastelloy X has been observed. It has been proposed that structural differences between both alloys determine the type of occurring radiation-induced processes. Excess energy deposited into materials' structure during ion-irradiation can lower the temperature of nucleation of high-temperature phases, which initiates the formation of grain boundary serrations.

## 1. Introduction

Nickel-based superalloys display a unique set of properties. They demonstrate excellent creep strength, superior corrosion resistance, and outstanding mechanical properties at a wide temperature range [1,2]. These materials can withstand long-term exploitation in load-bearing applications above 85% of homologous temperature [3]. Due to their high-performance ability, Ni-alloys are widely used in the most demanding applications, such as space and aerospace industries (i.e., turbine and engine blades or combustion chambers [3]). They are also considered as structural parts in elements of IV gen. nuclear reactor (i.e., intermediate heat exchanger [4], core barrel, or core support [5]). Such an exploitation environment also includes interaction with high radiation fluxes. It is known that radiation introduces numerous defects into crystal structures that impact the functional properties of materials [6]. Moreover, the material's response to irradiation strongly depends on the microstructural features of alloys. The interstitial and vacancy mobility are crucial factors of microstructure evolution during radiation damage. It has been proved that strengthening γ' particles in Ni-based superalloys dissolve as a result of radiation disordering [7,8]. Moreover, high levels of radiation damage result in more complex effects, such as voids formation and radiation-induced segregation of solute atoms, that may promote the deterioration of corrosion resistance [9,10].

However, purely solid-solution and carbide-strengthened alloys are less sensitive to radiation-induced precipitation dissolution effects [8]. These materials exhibit superior thermal stability for extended operating time and good ductility [11]. Nevertheless, contrary to dual-phase Ni-superalloys, purely solid-solution and carbide-strengthened materials exhibit lower creep resistance. Since grain boundary morphology and damage accumulation in their close proximity strongly influence the ultimate creep life, an alternative approach of grain boundary engineering offers an opportunity to improve the creep resistance. Employing an appropriate, precisely controlled heat treatment may trigger the generation of grain boundary serrations (GBS), which significantly alter deformation behavior [12,13]. The well-developed local architecture of grain boundary can delay crack propagation [13–15], significantly increasing creep life or delaying the onset of tertiary creep [15,16]. Furthermore, serrated grain boundaries elongate the diffusion path, which beneficially acts for oxidation-assisted cracking [8,15]. For this reason, the formation of GBS is an industrially very attractive phenomenon.

The transport of solute atoms and preferential growth of coherent γ' precipitates and carbides are deemed to assist the formation of grain boundary serrations [8,13,17,18]. Oriented growth of precipitations along grain boundary, when specific segments approach {111} low-index planes, provide the minimum interfacial energy between lattice and grain boundary. It must be highlighted that GBS may also be generated in the absence of carbides or γ' particles [19]. Thus, it was suggested that GBS might be promoted by diffusion and segregation of alloying elements in the vicinity of grain boundary [18,19]. It has been proved that irradiation enhances the redistribution of solute and impurity elements in the structure, leading to their enrichment in regions of dislocations and grain boundaries [20]. Furthermore, during irradiation damaging number of vacancies or interstitials in a metal's structure constantly increases [9]. It is known that the grain boundaries, dislocations, and phase interfaces act as defect sinks, where produced vacancies and interstitials may be annihilated by mutual recombination [21]. For these reasons, processes promoted by radiation-enhanced diffusion begin in close vicinity of structural defects.

As mentioned above, the exploitation environment in aerospace or IV gen. nuclear reactors includes harmful effects of radiation interaction. While the long-term performance of these materials in terms of thermal stability, mechanical properties, and corrosion resistance is widely described in the literature, the effects of irradiation remain unclear. For this reason, in the frame of this study, two types of commercially available Ni-based alloys (Hastelloy X and Haynes 230) designed for high-temperature performance were compared. This work focuses mainly on investigating radiation resistance and the impact of radiation defects on the material's properties. Hence, mechanisms accompanying radiation damage and their impact on materials' structural features – especially in the terms of processes occurring in the proximity of the grain boundary areas - must be carefully described and understood.

## 2. Materials and methods

In this study, two types of commercially available Ni-based alloys were investigated. Hastelloy X and Haynes 230 have been provided by Haynes International Co. [9]. Both materials were delivered in the shape of plates with thicknesses of 1.27mm and 1.02mm, respectively. According to the manufacturer, plates were hot rolled and annealed at 1149°C - 1177°C for Hastelloy X and 1177°C - 1246°C for Haynes 230. The chemical composition of both materials is presented in Tab. 1. Properties of both materials meet the requirements of Aerospace Materials Standards (AMS) – AMS 5536 [22] for Hastelloy X and AMS 5878 [23] for Haynes 230. Rectangular specimens with the geometry 15 x 15 mm and the thickness of the sheet were cut using Wire Electric Discharge (WEDM) technique. This methodology does not introduce additional stresses related to conventional machining. For further investigation, a high-quality mirror-finished surface was prepared by initial grinding with sandpapers (up to #1000) and subsequent electro-polishing at 10°C. The mixtures of perchloric acid with methanol (for Hastelloy X) or ethanol (for Haynes 230) were used.

**Table 1**. Chemical composition (in %$_{wt.}$) of research materials.

| Material | Cr | Fe | Co | Mo | Mn | W | C | Al | B | Si | P | Ni |
|---|---|---|---|---|---|---|---|---|---|---|---|---|
| Hastelloy X | 21.27 | 18.83 | 1.22 | 8.29 | 0.64 | 0.52 | 0.07 | 0.11 | <0.002 | 0.24 | 0.015 | Bal. |
| Haynes 230 | 21.87 | 1.23 | 0.20 | 1.46 | 0.50 | 14.24 | 0.10 | 0.37 | 0.04 | 0.31 | 0.07 | Bal. |

Specimens have been submitted to Ar-ion irradiation with an energy of 320 keV at 400°C. Both studied materials were irradiated with $1\times10^{15}$ and $1\times10^{16}$ ions/cms$^2$, corresponding to 1.2 dpa and 12 dpa, respectively. An additional set of specimens with covered surfaces was placed in the ion accelerator serving as reference materials in the later studies. Implementation of this procedure allowed for the separation effects of the temperature from the impact of radiation damage. The damage profile (Figure 1) was estimated by the NRT model employing the SRIM program [24]. Further details on the NRT model can be found in works [6,25]. For SRIM calculations, values of threshold energy (specific to each material element) were used as reported in Konobeyev's work [26].

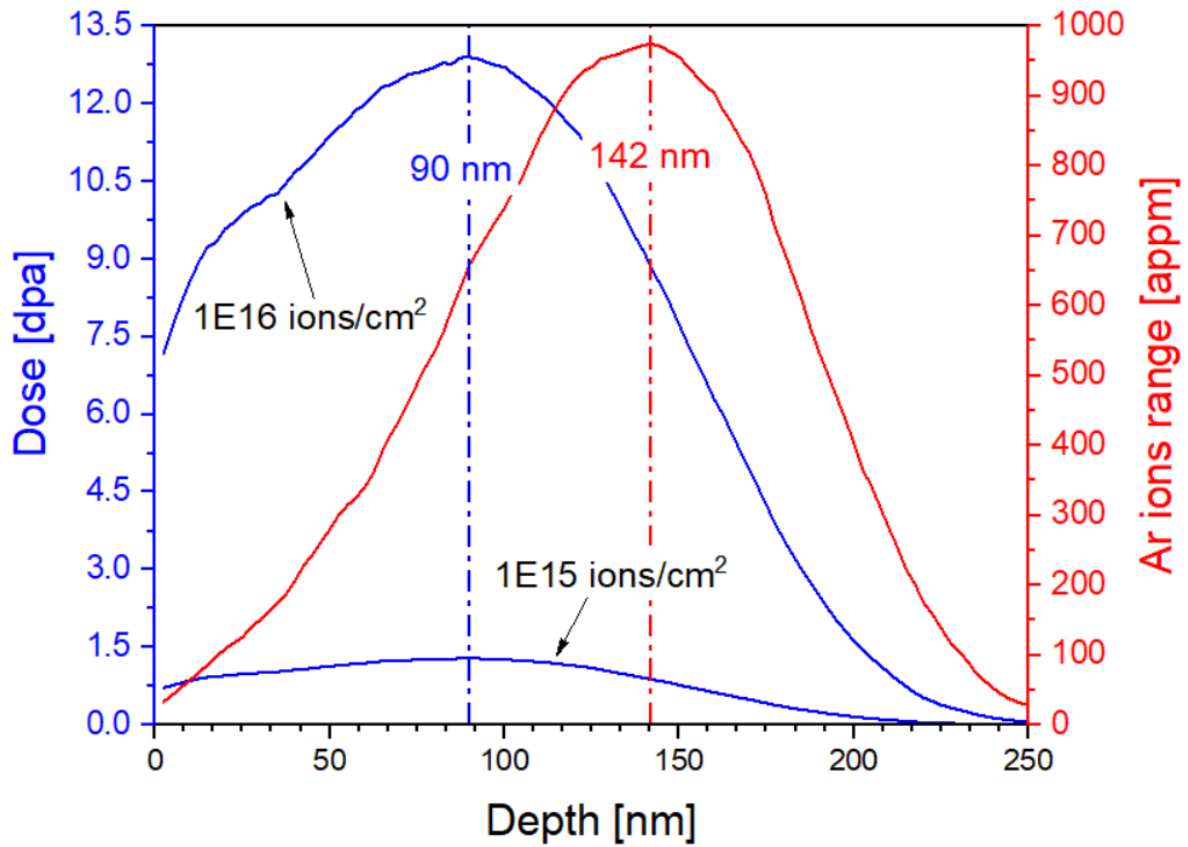

**Figure 1.** Damage accumulation (left blue axis) and irradiated Ar concentration (right red axis) in Ni-based alloys as a function of depth calculated by SRIM [24] with the NRT model. The depth positions of peak values at 90 and 142 nm are also marked with dashed lines separately.

The structural investigation was carried out by using Scanning Electron Microscopy (SEM) (provided by ThermoFisher Scientific - model Helios 5 UX), equipped with Electron Backscatter Diffraction (EBSD) and Energy Dispersive X-ray Spectroscopy (EDS) detectors. The Focused Ion Beam (FIB) lift-out technique was used to obtain electron-transparent lamellas from ion-irradiated zones. To minimize the impact of the $Ga^+$ ions on the microstructure of the materials, the final thinning was performed using a 2 kV beam. Detailed structure analysis has been performed using Transmission Electron Microscopy (TEM) (model JEOL JEM 1200EX II) operating at 120kV. The dislocation density in the ion-modified layer and as-received material has been estimated by applying the line intercept method described in [27]. Specimen thickness was determined with use of convergent beam electron diffraction by measuring Kossel-Moellensted fringes spacing.

Mechanical properties have been determined by using the nanoindentation technique. Experiments were performed on a NanoTest Vantage System provided by Micro Materials Ltd. One must remember that the tip shape is a critical factor during low-load nanoindentation [28]. Thus, the indenter tip's Diamond Area Function (DAF) was calculated for each indentation depth and used during the analysis. Calibrations were performed using Fused Silica (FS) as a standard material with defined mechanical properties. Tests were conducted at room temperature with a Berkovich-shaped diamond indenter (Synton-MDP). Indentations were performed in single-force mode using 12 forces ranging from 0.25 mN to 5.00 mN. Each measurement was repeated at least 20 times with 20 μm spacing between the indents. Characteristic parameters such as loading, unloading, and dwell time were set for 5s, 3s, and 2s, respectively. Oliver and Pharr method [29] was applied to extract nanomechanical properties from load-displacement (L-D) curves. In order to reduce the influence of bulk material, the indentation depth cannot exceed 10% of the investigated layer thickness [30]. For this reason, the estimated depth of collecting materials' response during described experiments was 20-200nm and included the area of interest – peak damage (Figure 1).

### 3. Results

Figures 2 and 3 show a general view of the studied materials' microstructure. In both alloys, numerous twin boundaries and precipitates are well visible. The EBSD analysis, as depicted in Figure 4, confirmed the presence of the FCC-matrix (γ phase) with the regular shape of grains. The orientation of grains in each material is random, thus no structural texture has been recorded. As highlighted before, numerous twin grains were observed. The twin boundary estimated fraction in both alloys is approximately 44% (Figure 4). This structure is typical for materials submitted to heavy deformation (hot rolling) and annealing. It is known that such treatment improves the material's mechanical properties [31,32].

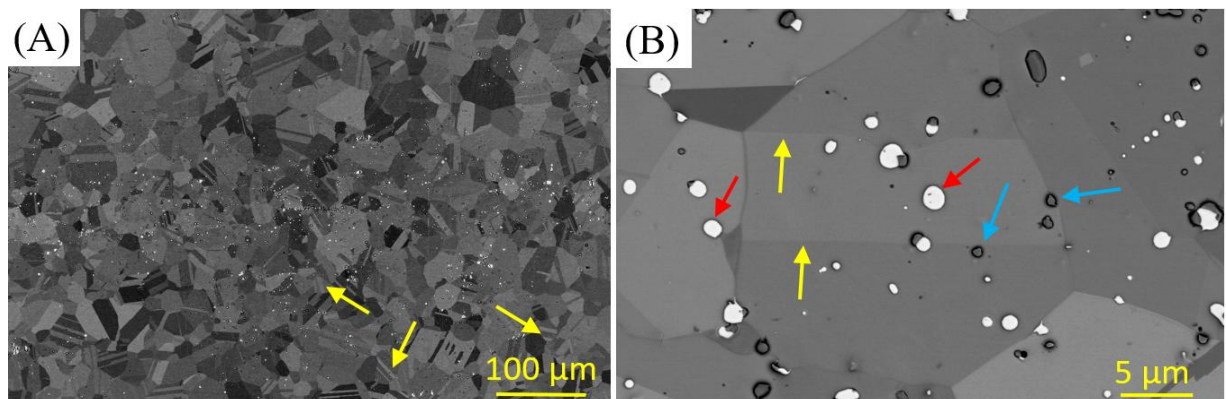

**Figure 2.** SEM images of Hastelloy X microstructure: (A) low-magnification general view of microstructure, and (B) focus on grain and precipitates. Numerous twin boundaries (indicated by yellow arrows), and two types of particles (coarser – probably $M_6C$ type – pointed by red arrows, and finer probably $M_{23}C_6$ type – indicated by blue arrows) are visible. Images were taken using Circular Backscatter Detector (CBS).

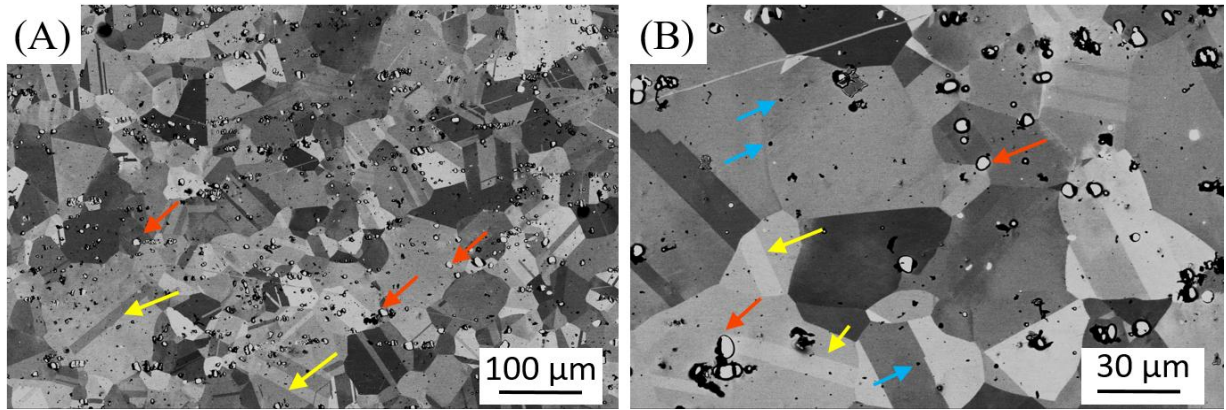

**Figure 3.** SEM images of Haynes 230 microstructure: (A) low-magnification general view of microstructure, and (B) focus on grain and precipitates. Numerous twin boundaries (indicated by yellow arrows), and two types of particles (coarser $M_6C$ type – pointed by red arrows, and finer $M_{23}C_6$ type – indicated by blue arrows) are visible. Images were taken using Circular Backscatter Detector (CBS).

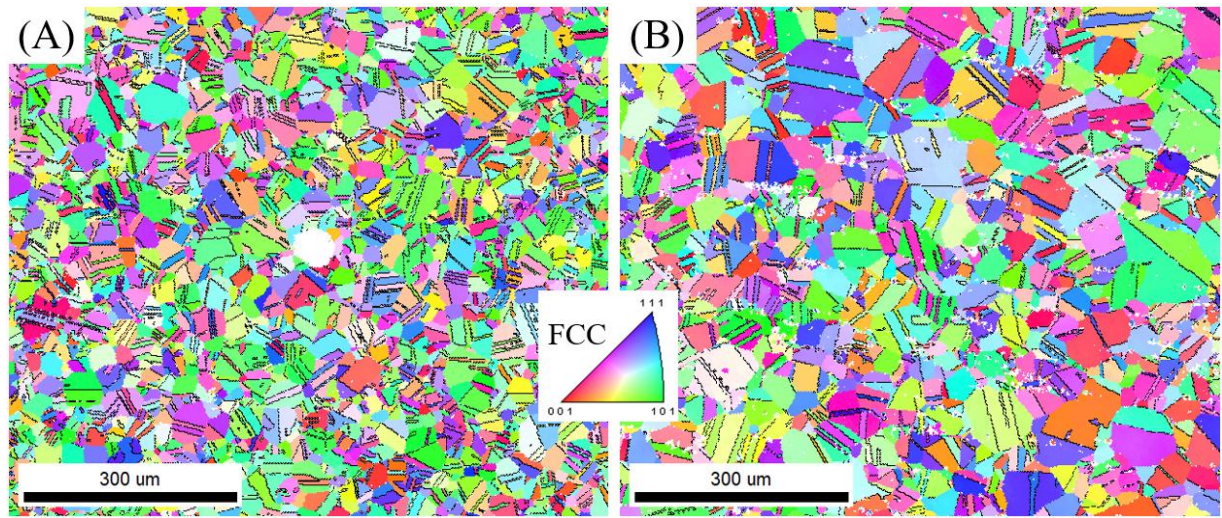

**Figure 4.** EBSD orientation maps of (A) Hastelloy X, and (B) Haynes 230 specimens. Twin boundaries are marked by black lines.

Furthermore, the EBSD analysis allowed the determination of grain size in both alloys (Figure 5). One may notice that Hastelloy X is characterized by a slightly finer grain size when compared to Haynes 230. As a result, the calculated average grain size of Haynes 230 is 35.4 µm, while the average grain size of Hastelloy X equals 22.6 µm.

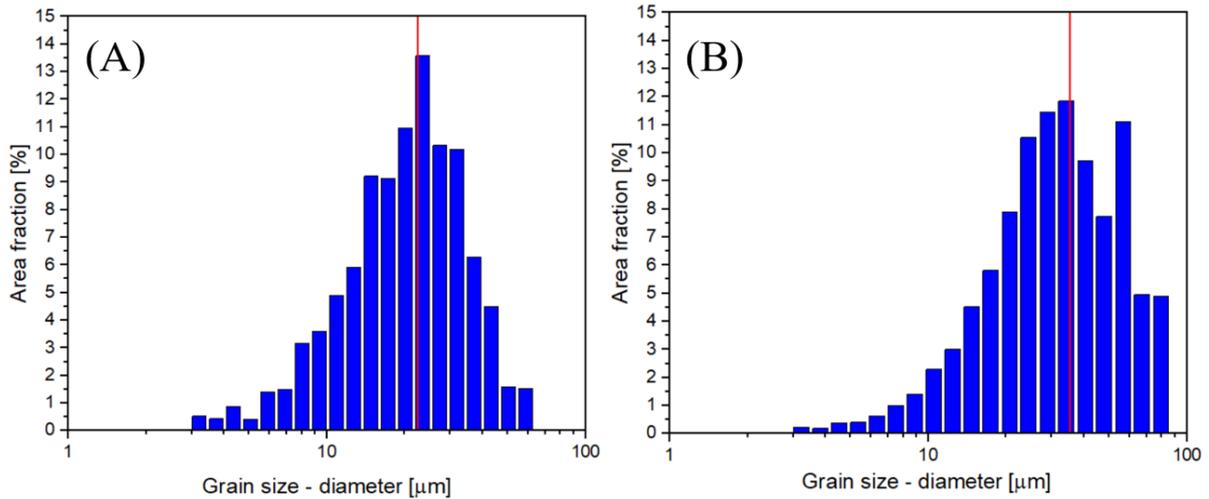

**Figure 5.** Grain size histogram for (A) Hastelloy X, and (B) Haynes 230 specimens. Mean values are marked by red lines.

In both materials, two populations of precipitates (finer and coarser) may be found (see Figures 2 and 3). The EDS analysis has shown that precipitates differ in chemical composition (Figure 6). One group of precipitates exhibits observable enrichment in Cr, whereas the second group may be identified as particles with a high concentration of Mo. According to literature data [33–35], these particles may be identified as $M_{23}C_6$ precipitates. The coarser carbides are probably $M_6C$-type precipitates rich in Cr.

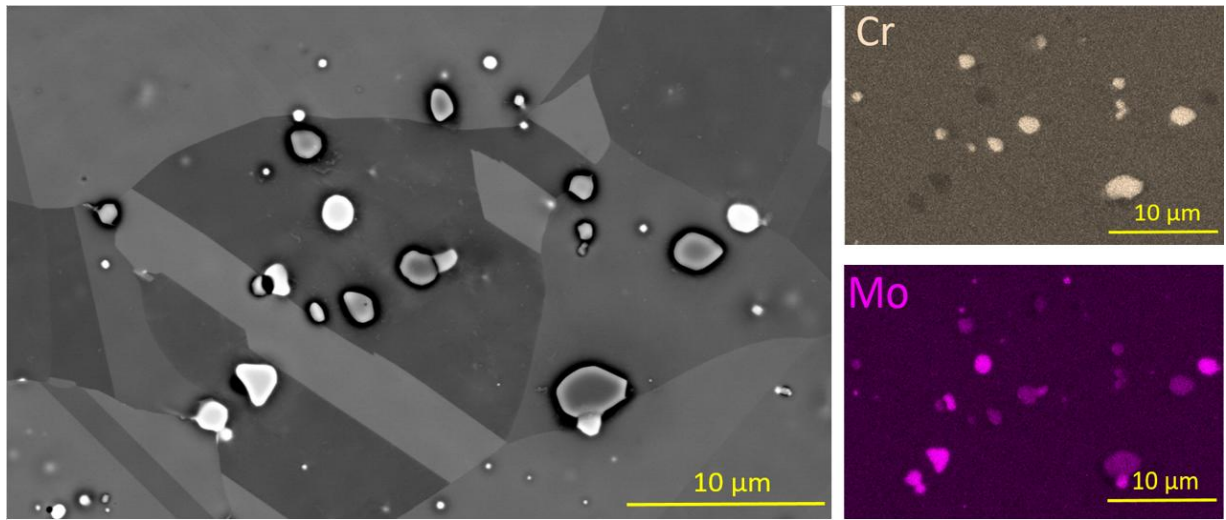

**Figure 6.** SEM images of Hastelloy X carbides with corresponding EDS maps of elements: Cr and Mo.

In order to confirm the structure of observed particles, electron-transparent lamellas have been prepared by FIB lift out-technique. The STEM observations showed that two carbide types (Figure 7) co-exist in the investigated material. The significant differences in contrast (Figures 7A and 7B) suggest essential differences in particle chemical composition – as observed during SEM-EDS investigations (Figure 6). Chemical composition has been confirmed by STEM-EDS analysis, which revealed that dark particles (no. 1 and no. 3 in Figure 7A) are rich in Cr and Mn, whereas carbide marked as no. 2 (Figure 7A) is enriched in Mo and W (Figure 7C).

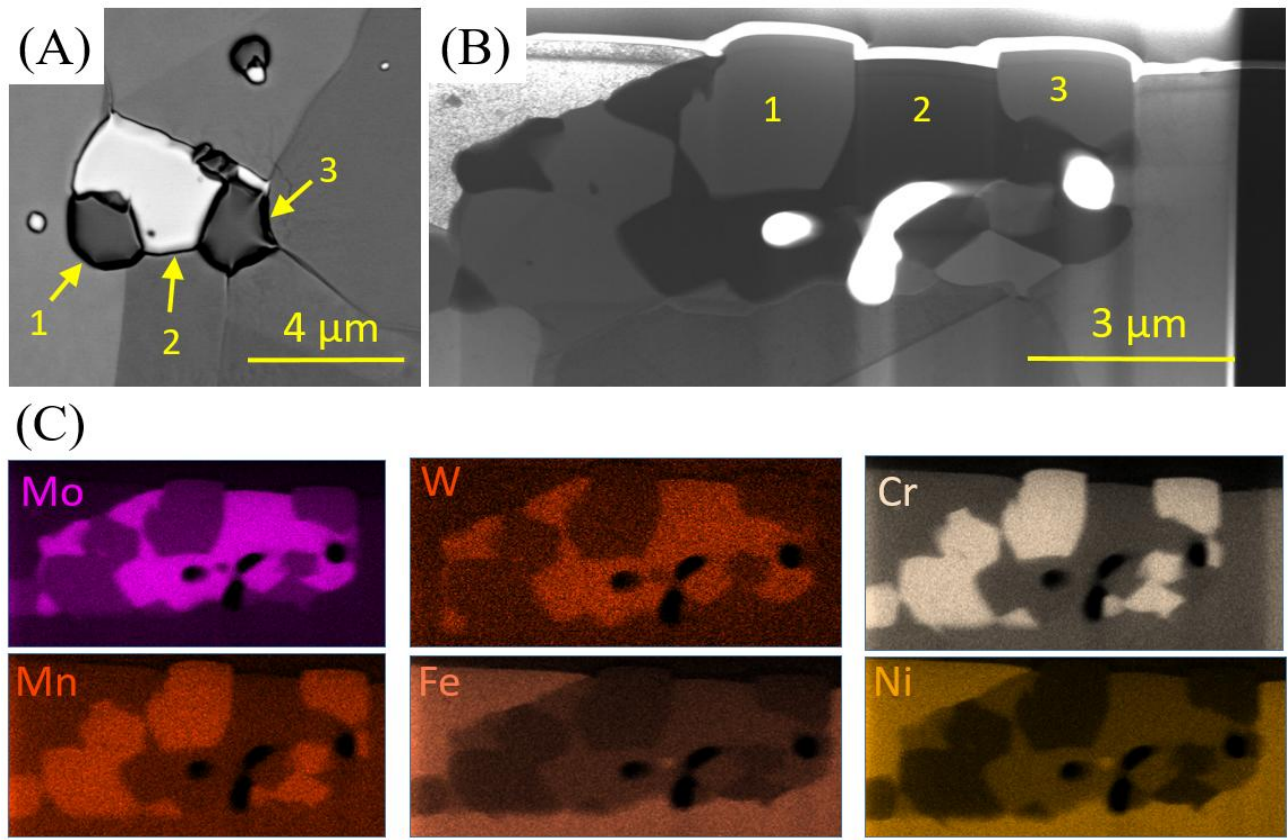

**Figure 7.** Images of precipitates in Hastelloy X microstructure: (A) SEM image of three co-existing carbides; (B) STEM image of the cross-section of particles observed in figure (A); (C) STEM-EDS maps of carbides observed in figure (B).

As a next step, an in-depth analysis of Selected Area Electron Diffraction (SAED) patterns obtained from observed two types of particles (visible in Figure 7) has been performed and presented in Figure 8. It has been confirmed that two types of carbides are observed: $M_6C$ (marked as no. 1 in Figure 8) with lattice parameter *a* of 11.23 Å, and $M_{23}C_6$-type of particles (marked as no. 2 in Figure 8) with lattice parameter *a* of 10.86 Å. The calculated lattice parameters are in full agreement with the literature data [36–40]. According to obtained STEM-EDS maps (Figure 7), $M_6C$ particles are enriched in Cr and Mn (particle no. 1 in Figure 8), while $M_{23}C_6$ precipitates are carbides of Mo and W (particle no. 2 in Figure 8). Both types of carbides are characterized by FCC structure. Similar results may be found in the literature [33–35].

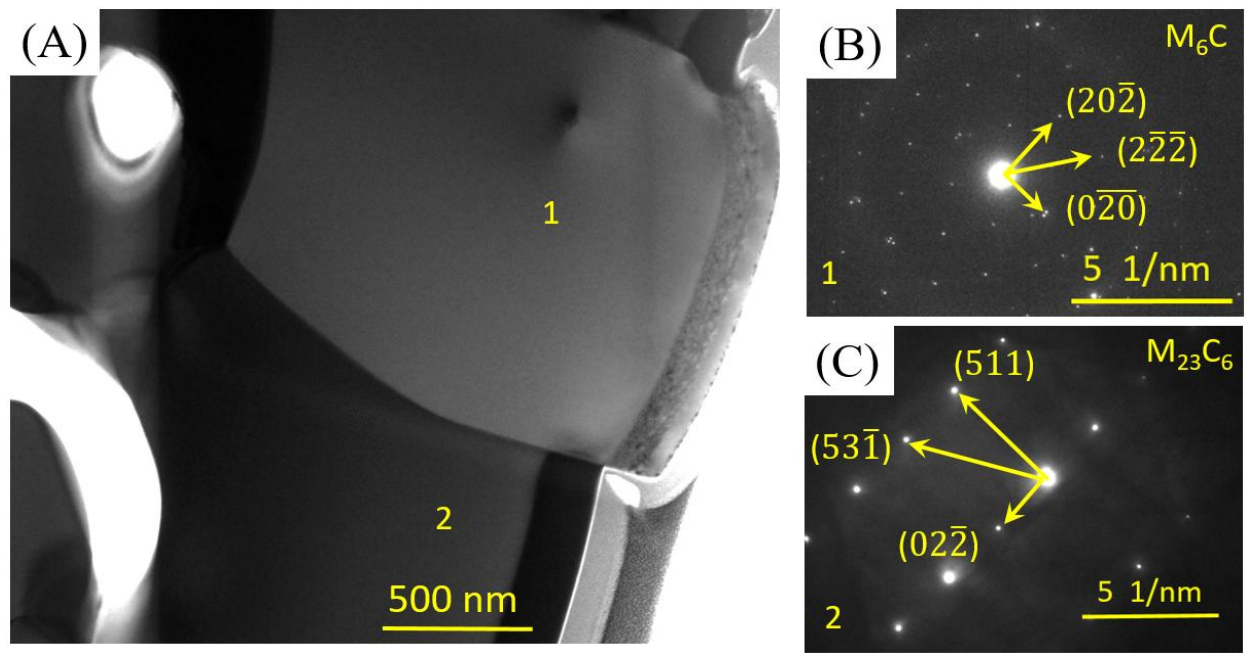

**Figure 8.** TEM images of Hastelloy X carbides: (A) general view of identified carbides no. 1 and no. 2; (B) SAED pattern of carbide no. 1, and (B) SAED pattern of carbide no. 2.

To determine the effects of ion-irradiation on the materials' microstructure, TEM observations of lift-out lamellas prepared through $Ar^+$-modified layers were performed. Detailed analysis of modified layers (Figure 9 and Figure 10) revealed the presence of numerous radiation-induced defects. As expected, their distribution is non-uniform. The density of dislocations increases continuously, approaching its maximum value, visible as a thick dark interlayer (pointed by yellow arrows in Figures 9 and 10). This area corresponds to the peak damage localized at a depth of approx. 110 nm for Hastelloy X (Figure 9) and 140 nm for Haynes 230 (Figure 10). Below this boundary, the density of dislocations decreases rapidly. Such distribution of the defects is strictly related to the inhomogeneous dose profile (Figure 1), which is related to the inhibition of knocked-out atoms in the material's crystal lattice. One may notice that the location of the peak damage determined during TEM observation is slightly deeper than the depth predicted by simulations (Figure 1). This mismatch is related to two factors: (i) the difference between assumptions of model alloy and real material and (ii) the temperature of the experiment. In contrast to SRIM assumptions, real experiments include natural fluctuations of local chemical composition and structure of materials and some variation of radiation rate and its fluency. Although, in this case, the irradiation temperature is the main factor influencing the depth of peak damage. It is known that an increase of temperature to 0.35 of melting temperature ($T_m$) leads to intensified mobility of radiation defects [41]. In this case, the irradiation temperature is 400°C, which is very close to the above-mentioned limit for studied materials. The temperature during the experiment was measured by the controlling temperature of bulk material, not the irradiated surface. Hence, in some areas submitted to irradiation, the limit of $0.35T_m$ may have been exceeded. This additional energy enhances the diffusion of defects. Increased mobility of defects at the temperature of the experiment and their gradient related to the direction of irradiation flux determine the path of defects' flow – deeper from the surface. However, the observed difference is very small and may be irrelevant to this study.

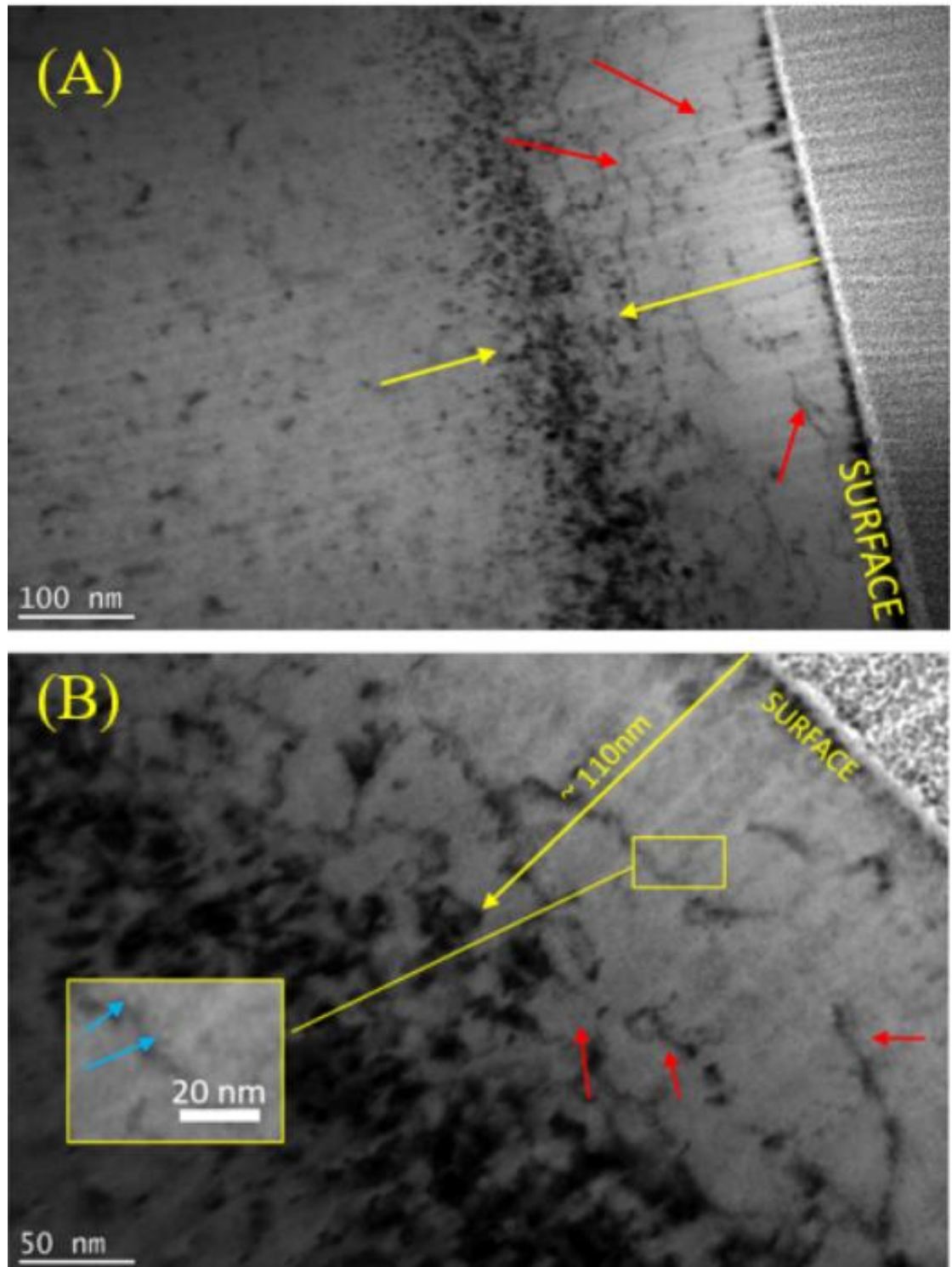

**Figure 9.** TEM images of Hastelloy X submitted to $Ar^+$ irradiation up to $1 \times 10^{16}$ ions/cm$^2$: (A) general view of irradiated affected layer, and (B) under focus magnification image with visible Ar bubbles (white dots indicated by blue arrows in yellow rectangle magnification). Red arrows indicate crystal defects, and yellow arrows point to the area at peak damage.

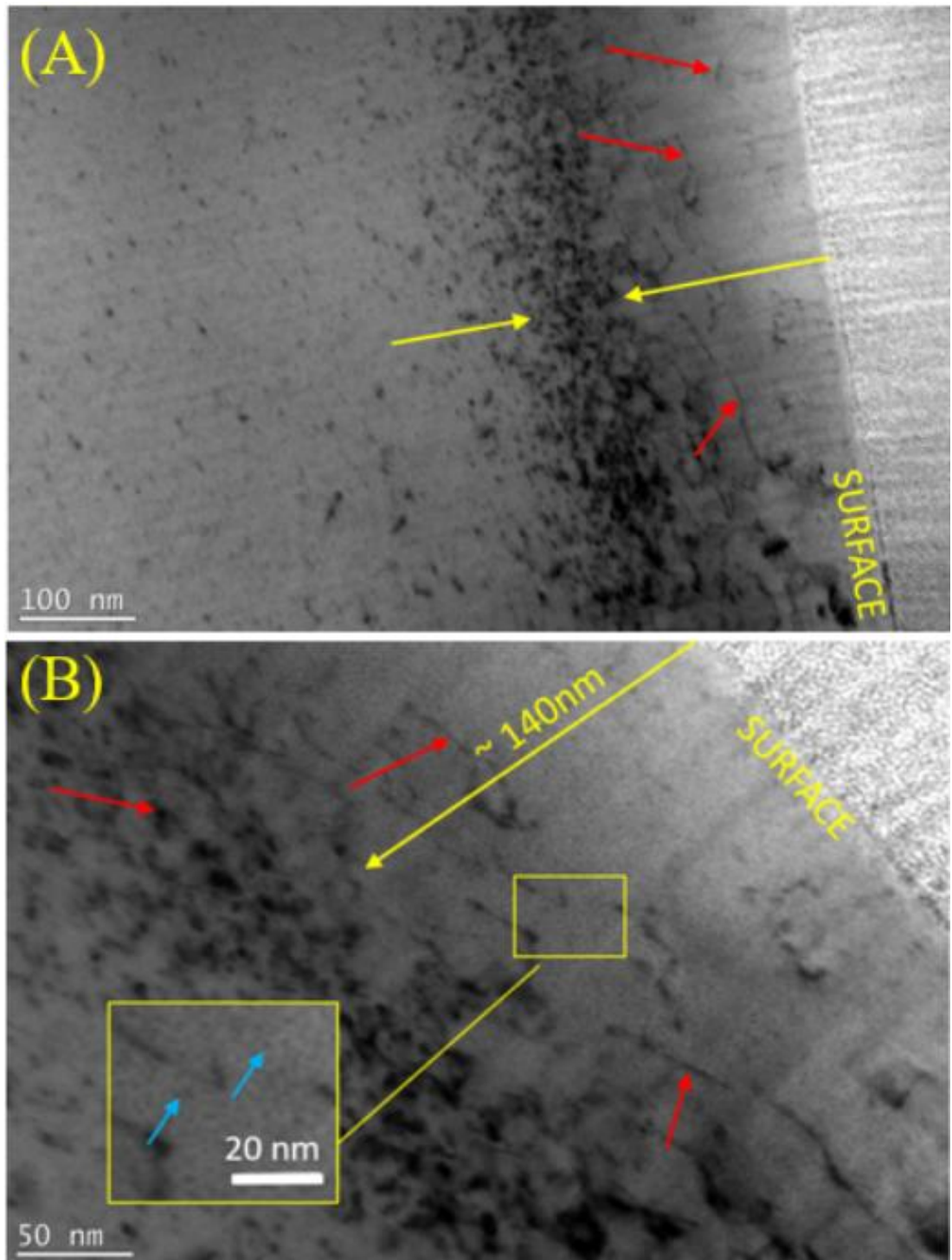

**Figure 10.** TEM images of Haynes 230 submitted to Ar$^+$ irradiation up to $1 \times 10^{16}$ ions/cm$^2$: (A) general view of irradiated affected layer and (B) over-focus magnification image with visible Ar bubbles (black dots indicated by blue arrows in yellow rectangle magnification). Red arrows indicate crystal defects (dislocations and point defects), and yellow arrows point to the area at peak damage.

Detailed examination of the ion-modified layer reveals the presence of dislocations in the form of short spontaneously arranged lines and other defects visible as black dots. As presented in the work [27], the implementation of the line intercept method provides a quantitative analysis of irradiation effects, as shown in Table 2. One can notice that the ion-irradiation increases dislocation density, especially in Haynes 230. In this material, dislocation density increases almost twice as an effect of irradiation. For Hastelloy X increase of over 35% is observed. Moreover, the change in contrast of images taken under focus (Figure 9B – white dots indicated by blue arrows) and over focus conditions (Figure 10B – black dots indicated by blue arrows) has confirmed the presence of nanometric voids homogenously distributed in the modified layer. Observed features indicate the presence of Ar bubbles trapped in the microstructure.

**Table 2.** Estimated density of dislocations in the studied materials.

| | Dislocations density [$1/\mu m^2$] | |
|---|---|---|
| **Material** | Virgin | Ion-irradiated ($Ar^+$, fluence $1 \times 10^{16}$ ions/cm$^2$) |
| **Hastelloy X** | 3.54 x 10$^{14}$ | 4.81 x 10$^{14}$ |
| **Haynes 230** | 2.65 x 10$^{14}$ | 5.18 x 10$^{14}$ |

The nanoindentation technique was used to evaluate changes in mechanical properties developed as ion irradiation effect (Figure 11). Increased hardness at a low force range is well visible in studied cases. Applying higher forces results in a rapid hardness decrease. This phenomenon is known as Indentation Size Effect (ISE) and is well described in the literature [42–44]. It is related to the amount of the volume of the material displaced by the indenter. Using lower loads during tests leads to the activation of geometrically necessary dislocations nearby and the activation of the nearest slip systems (Figure 12A). Experiments conducted using higher loads result in a more significant material volume recording response, including obstacles for dislocation movement (such as grain boundaries or precipitates) and slip systems exceeding nearest proximity (Figure 12B). Since conducted irradiation procedure allows for the modification of a very thin layer (Figure 1), revealing the effects of this experiment requires analysis of the results of extremely shallow nanoindentation. In this case, the study must focus on results obtained by applying the lowest forces (0.25 or 0.5 mN). According to the presented results, indentation with such loads leads to penetration of materials at a depth of approx. 25-40 nm. Thus, it must be highlighted that the ISE phenomena intensify observed irradiation-induced changes. Furthermore, due to the fact that the plastic zone below the indent is 7-10x deeper than the penetration depth [42,44], the impact of unmodified material on results cannot be neglected. Nevertheless, despite the above-mentioned detrimental factors, the influence of irradiation on mechanical properties is still well visible.

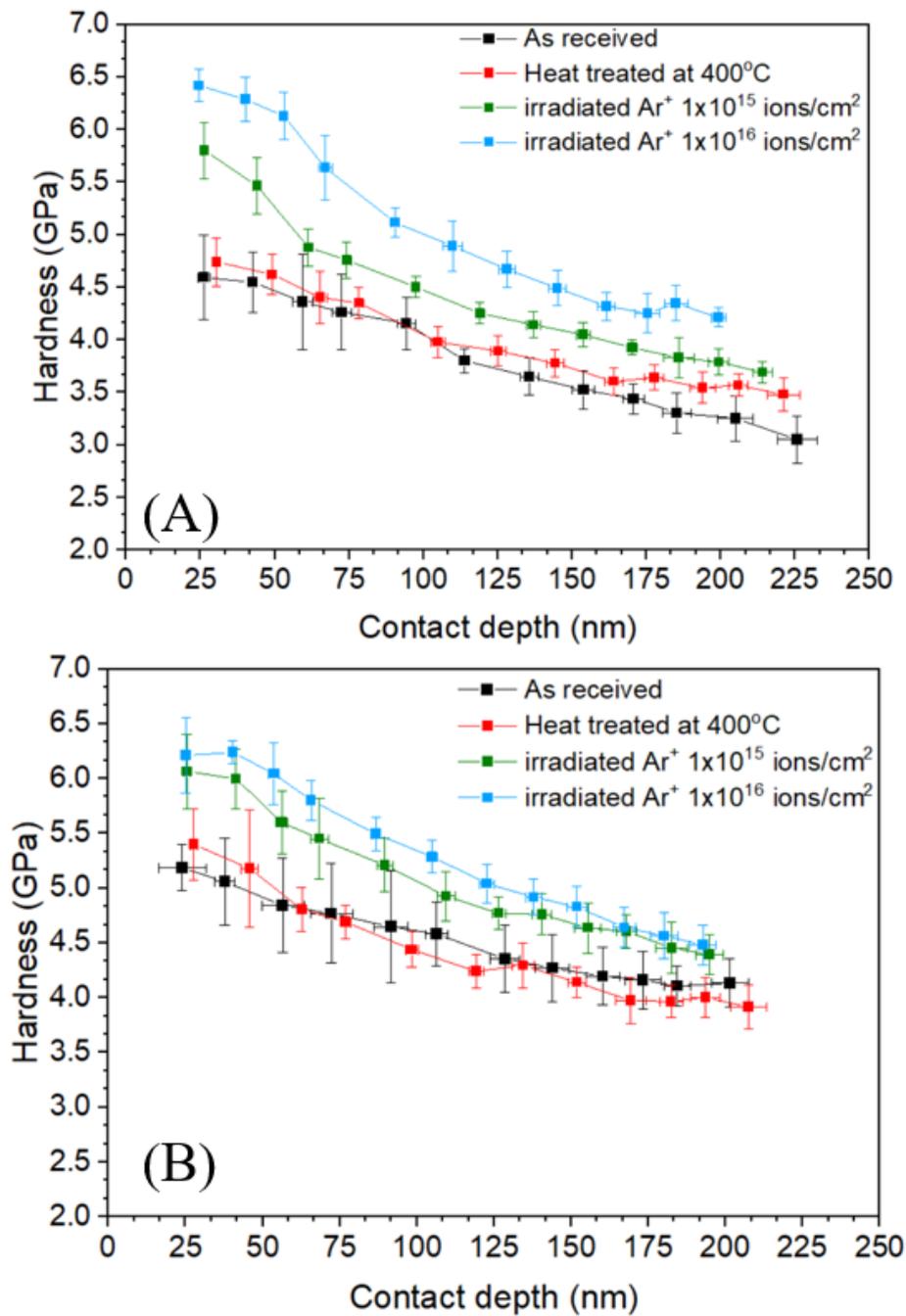

**Figure 11.** Nanohardness versus plastic indentation depth of specimens submitted to ion-irradiation at 400ºC: (A) Hastelloy X, and (B) Haynes 230.

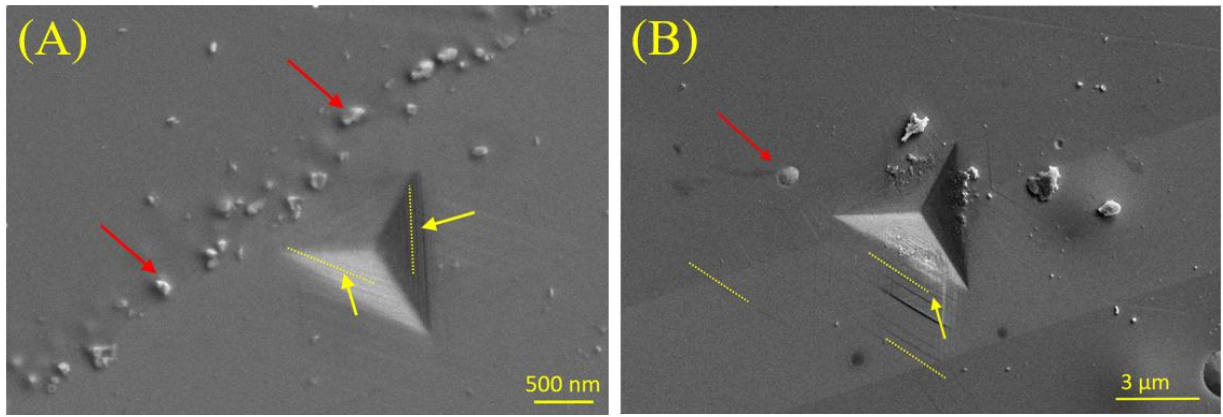

**Figure 12.** SEM images of the indents after tests performed with (A) low and (B) high load recorded on Haynes 230. The red arrow indicates the carbide's presence, whereas the yellow arrows point to activated slip traces. The direction of slip systems is marked by dashed lines (to guide the eye).

The hardness decreases asymptotically to the value of the as-received material with increasing contact depth (Figure 11). Moreover, Ar-irradiation leads to the continuous increment of hardness – the higher fluence of Ar ions introduced to the structure, the higher the nanohardness value is recorded. This phenomenon is known as a radiation hardening effect and is related to generating structural defects in material [45–47], as confirmed by TEM observations. The most intensive hardness increase is visible at the 25-50nm depth range. Probing this region is strictly related to the response of the ion-irradiated layer. It must be highlighted that investigated irradiation hardening effect is partially "covered" with the described ISE effect.

When comparing results obtained for as-received materials, one may notice that Haynes 230 displays higher values of nanohardness than Hastelloy X (Figure 11). Heat treatment causes a slight increase in the mechanical properties, although observed changes are below 5% of the original values for both investigated materials (Figure 11 and Figure 13). For this reason, the influence of the temperature may be regarded as negligible. Since these materials are designed for temperature applications exceeding 400ºC (even two times higher than irradiation temperature), this effect is not surprising.

A comparison of mechanical parameters has revealed that Hastelloy X displays a much more intensive hardening effect than Haynes 230 (Figure 13). A 25% increase in the initial hardness of Hastelloy X due to irradiation up to 1.2 dpa has been recorded. At the same time, the properties of Haynes 230 have changed by 17%. Further increase of the damage up to 12 dpa leads to nanohardness increase by 40% for Hastelloy X. Radiation hardening of Haynes 230 is twice as lower as Hastelloy X.

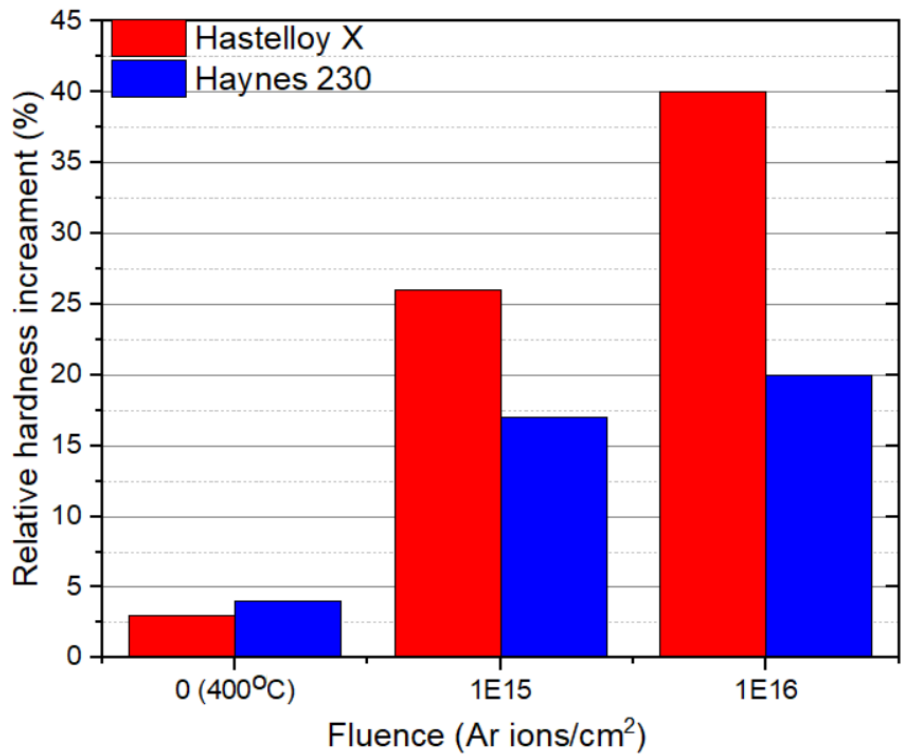

**Figure 13.** Relative hardness increase of the investigated materials as a function of ion fluence.

The further study focused on changes introduced due to ion-irradiation on other structure elements – carbides and grain boundaries. SEM images of carbides in Hastelloy X (Figure 14) suggest that particles slightly changed shape due to irradiation. The irradiation process leads to the evolution of short "arms" originating from carbides. The similar geometry of carbides is hardly visible in the material before Ar-irradiation (Figure 14A). Carbides present in as-received material may be characterized as regular and oval. Comparing these images with images taken on Haynes 230 (Figure 15), one may notice that such alteration of carbides' morphology as a result of irradiation is hardly recorded.

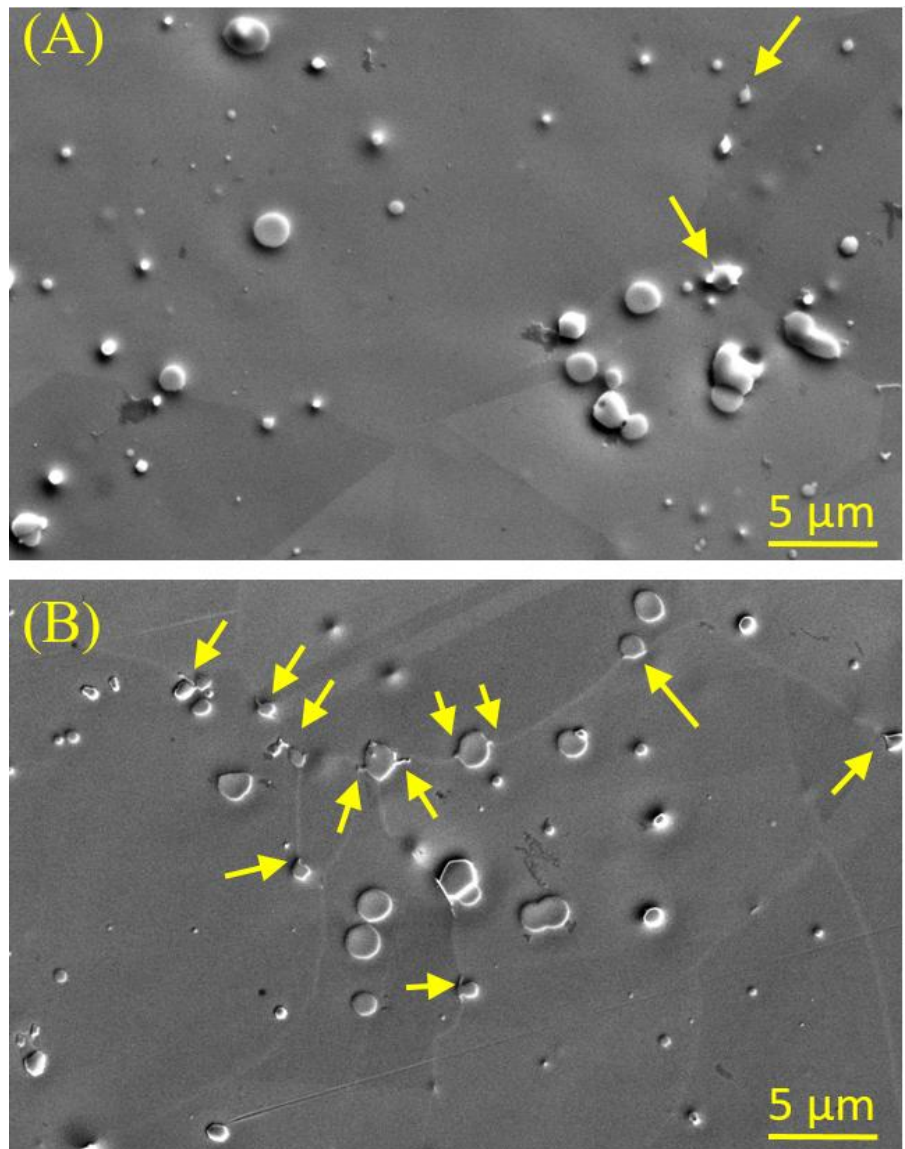

**Figure 14.** SEM images of grain boundaries and surrounding carbides in Hastelloy X (A) in the as-received state, and (B) after ion irradiation up to $1 \times 10^{16}$ Ar-ions/cm$^2$. Yellow arrows point to "arms" of carbides localized at the grain boundary.

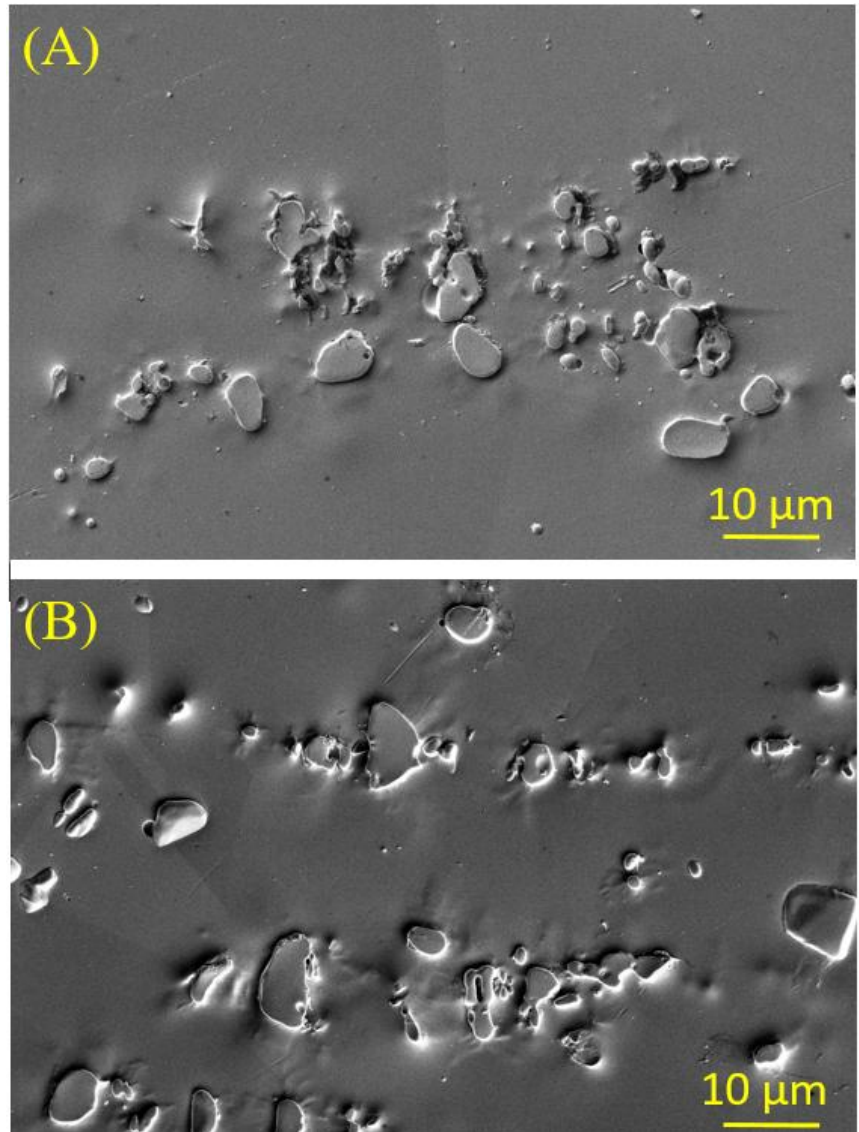

**Figure 15.** SEM images of grain boundaries and surrounding carbides in Haynes 230 (A) in as received state and (B) after ion irradiation up to $1\times10^{16}$ Ar-ions/cm$^2$.

It has been confirmed that two types of carbides (Figure 7 and Figure 8) may be found. For this reason, it is necessary to determine which type of carbide undergoes this particular morphology evolution, as presented in Figure 14. Thus, appropriate particles with "arms" have been chosen for detailed investigation, as shown in Figure 16. The analysis of SAED patterns has confirmed $M_6C$ - type structure, while STEM-EDS maps have revealed Mo and W-enrichment in each particle.

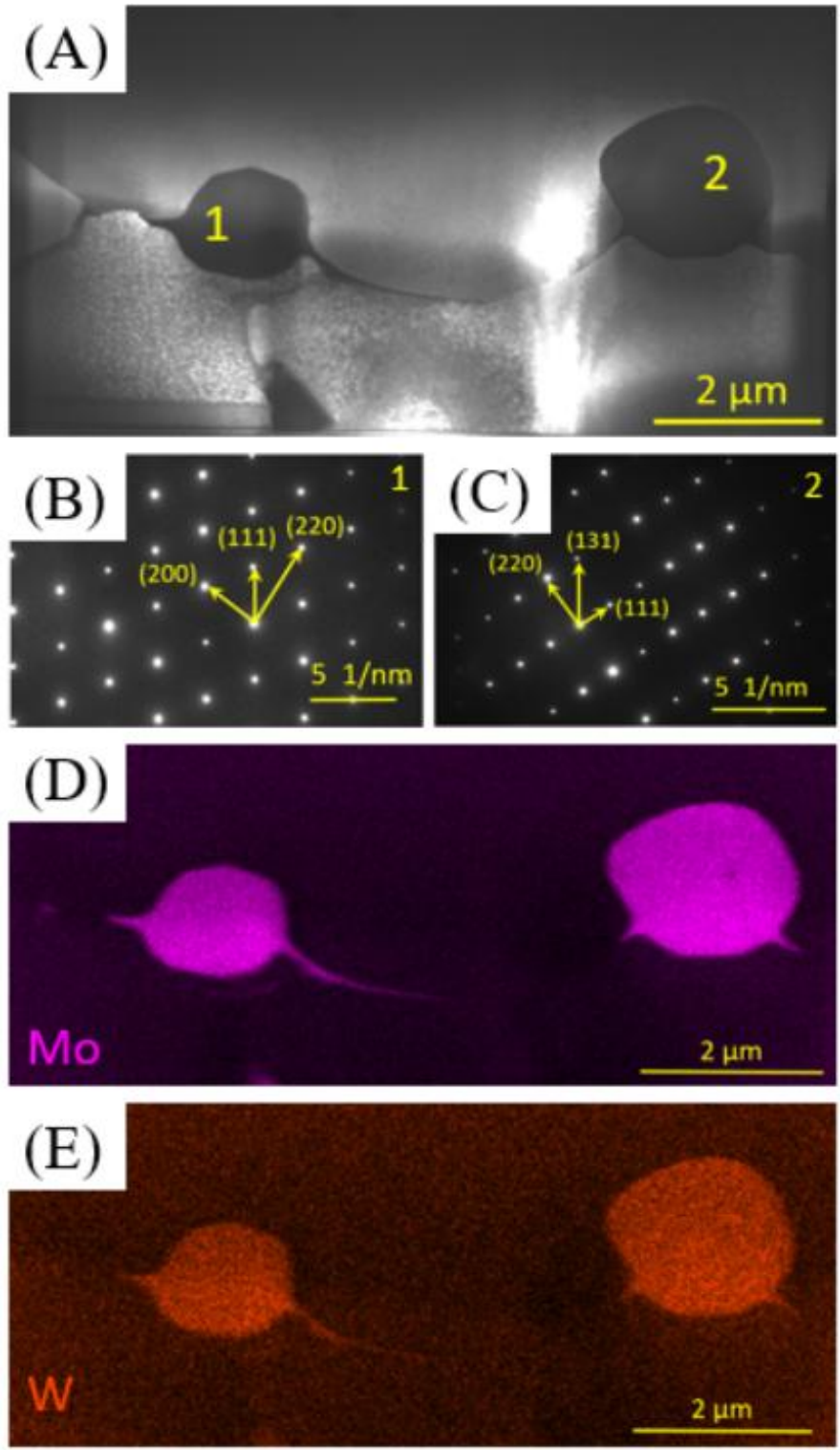

**Figure 16.** Images of the cross-section of carbides submitted to ion-irradiation process (A) STEM-micrograph with visible two carbides (marked as no. 1 and no. 2) localized through grain boundary with noticeable "arms"; (B) SAED pattern corresponding to carbide no. 1, (C) SAED pattern corresponding to carbide no. 2, (D) STEM-EDS map of Mo-enrichment, and (E) STEM-EDS map of W-enrichment.

## 4. Discussion

Structural investigation of studied materials has confirmed that both materials are characterized by FCC structures (Figure 4) and are strengthened by two types of carbides: $M_{23}C_6$ and $M_6C$ (Figure 7 and Figure 8). Both materials have smooth grain boundaries with numerous twins (Figure 2 - 4). Ion irradiation leads to visible radiation hardening effect that resulting in a much more intensive response of Hastelloy X than Haynes 230 (Figure 11 and 13). These effects are strictly related to the generation of dislocation and point defects during Ar-ions damaging, as confirmed by TEM imaging (Figure 9 and Figure 10). Detailed analysis of dislocation density (Table 2) reveals that the defect increase is much higher in Haynes 230 than in Hatestelloy X. Reported result may be slightly surprising, considering that Hastelloy X exhibits a much higher hardening effect than Haynes 230. Combining this observations with nanoindentation results (Figure 11) suggests more complex strengthening mechanisms than those related strictly to the number of generated radiation defects. This could be associated with the observed carbides' morphology in both materials. It has been proved that in Ni superalloys effect known as Grain Boundaries Serrations (GBS) is present (Figure 17) [13,48,49]. The mechanism accompanied by the generation of GBS is associated with a heterogeneous nucleation of the strengthening γ' phase at the grain boundaries [48]. Furthermore, it has been proved that carbides localized near the grain boundaries play a crucial role in generating GBS [13,50]. Although, it must be highlighted that serrations may be obtained by appropriate heat treatment at a temperature range of 1000-1300ºC and combined with a specially controlled cooling rate [13,48]. It is known that the GBS's presence extends Ni-based superalloys' exploitation time by providing meaningful improvement of creep strength [13,49] as well as beneficially influences corrosion resistance [13–15].

Images of carbides in Hastelloy X submitted to ion irradiation (Figure 14B) suggest that some very first stages of GBS evolution have occurred. Short "arms" aiming through grain boundaries and originating from the particles mentioned above are well visible. This morphology is similar to that observed in heat-treated Ni superalloys with present GBS (Figure 17) [27]. Although, the generation of serrated grain boundaries requires temperature high enough to promote precipitation process of the γ' phase and/or transformation of present carbides. Obtained results suggest the possibility of starting the very beginning stages of this process at a much lower temperature range – at least, processes preceding regular nucleation. Generation of a high number of defects during the irradiation process and increased mobility of defects at experiment temperature (400ºC) may provide sufficient energy to alter the morphology of carbides. This change may be necessary to promote following regular GBS evolution. Thus, the two effects of irradiation may be proposed and discussed:

(i) Excess energy deposited in the structure during the irradiation process combined with temperature-enhanced mobility of defects lead to intensification of diffusion processes, altering the carbide's structure near the grain boundaries into more stable forms. These processes may be considered as early-stages of GBS' evolution;

(ii) The ion irradiation process significantly decreases the temperature necessary for GBS evolution – at least for starting this process or accelerating its' kinetics. According to the literature, process of GBS formation includes transformation of $M_6C/M_{23}C_6$ carbides and heterogeneous nucleation of γ' phase at the grain boundaries [48,51]. For this reason, authors believe that in presented experiment, the evolution of precipitates at the proximity of grain boundaries is the mechanism preceding carbides' transformation and further γ' phase nucleation.

It must be highlighted that the both proposed effects require processes that take place at a much higher temperature regime than those used in the presented experiment. Nevertheless, it has been proved that nucleation of high-temperature phases in Ni-superalloys may occur below the regular temperature of their presence. For example, the σ phase may occur in Ni-based alloys due to long-term service or prolonged heat treatment [52]. Sakthivel *et al.* [53] demonstrated the existence of a high-temperature σ phase in Hastelloy X submitted to 550°C, which is a much lower temperature than expected from the literature. Since it has the same crystal structure as $M_{23}C_6$ (but without carbon atoms), the σ phase is believed to be a product of carbide transformation [52]. Moreover, Sakthivel *et al.* [53] also found coexisting two types of carbides (as in this work - Figure 8). They were spotted in the direct neighborhood in materials exposed to high temperatures. The same effect has been reported by Bai *et al.* [51]. Since $M_6C$ particles are unstable at high temperatures, these precipitates undergo progressive decomposition. Atoms released in this process contribute to the nucleation and growth of $M_{23}C_6$ carbides [51]. For this reasons, depending on the exploitation environment, partial decomposition of $M_6C$ carbides may occur at a lower temperature regime when the diffusion coefficient increase due to the presence of excess point defects – as in presented experiment. Then, the driving force of the $M_6C$ decomposition may be lowered and the process starts at lowered temperature. As a result, following nucleation and growth of $M_{23}C_6$ carbides may be expected, what is a crucial factor in generating fully developed GBS [13,50].

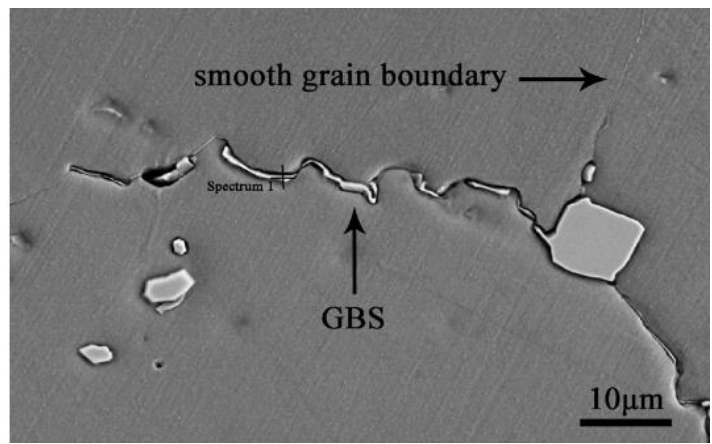

**Figure 17.** SEM images of smooth grain boundary and grain boundaries serrations (GBS). Image reprinted from work of Jiang *et al.* [13] with permission of Elsevier.

Ion irradiation is a process that damages the crystal structure by introducing material defects that usually evolve into more complex forms. In our experiment, energy deposited during ion irradiation combined with high temperature (which promotes defects mobility) are factors that influence the process of solute atoms reorganization. Authors believe that excess energy deposited in the Hastelloy X structure due to Ar-ion radiation is utilized to alter carbide's structure nearby the grain boundary. The observed change in carbide morphology (previously suggested in the literature [51,53]) indicates that the particle decomposition process has started. Conducted analysis of SAED patterns (Figure 16-18) confirms the $M_6C$ structure of carbides that have undergone a morphological alteration. According to the literature [51], transformation of $M_6C$ particles into $M_{23}C_6$ precipitates may be expected in the next stage. Our study shows that this phenomenon occurred at a lower temperature than reported in the literature. Presented results show that early stages of atoms' reorganization to promote high-temperature phase nucleation have been initiated as a result of energy deposition through ion-irradiation process. The observed morphology of the carbides in studied materials submitted to ion-irradiation and obtained SAED patterns suggest that serrated grain boundaries are about to form. Thus, according to literature data [51,53], one may expect that after carbides transformation in the proximity of grain boundaries, the γ' phase precipitation followed by GBS formation may take place.

Considering the proposed mechanisms, it is necessary to explain why similar effects are not recorded in Haynes 230 specimens. One may note that Hastelloy X is characterized by a finer structure than Haynes 230 (Figure 4 and 5). Moreover, the precipitations' size is much smaller in this material (Figure 2 - 3). These differences in as-received structure also determine the types of processes occurring during ion-irradiation. It is known that grain boundaries and precipitates act as defect sinks. Recombination of defects in grain boundaries and interfaces between particles and the matrix are fundamental mechanisms of reducing energy in material [54]. If these processes are insufficient, additional, more complex defects, such as dislocation lines or dislocation loops are generated.

Since Haynes 230 displays a much lower grain boundary density, the abovementioned point defect loss processes do not allow for efficient energy reduction. Thus, large areas inside the grains - mostly free from the dislocations - are utilized to generate new defects. In discussed case, this mechanism is energetically more favorable to rapid energy reduction than the reorganization of solute atoms and the transformation of structural elements (which consumes much more energy). Furthermore, the shape and size of the precipitates do not promote the processes described in the previous paragraph. Carbides in Haynes 230 are relatively coarse and distributed in blocks (Figure 3). Such morphology and distribution significantly reduce energy accumulated in the particle/matrix interface. Hence, the energy accumulated in as-received material and combined with energy deposited during ion irradiation is insufficient to initiate processes related to the reorganization of solute atoms and the beginning of decomposition of carbides. Thus, new dislocations are generated to reduce energy in areas that may be considered free from defects – in this case, the grains' interior. The differences between basic structural characteristics determine the type of mechanisms occurring during ion irradiation and the number of generated new defects. For this reason, the dislocation density increase as a result of ion-damage is much more intensive for Haynes 230 than for Hastelloy X (Table 2). The structural features of Haynes 230 do not reduce energy in the process of decomposition of carbides and nucleation of other phases. The opposite effects are observed in the Hastelloy X. In this material, basic mechanisms of energy reduction (recombination in the matrix, diffusion to defects sinks, and generation of new dislocations) are inefficient, and the system is characterized by energy sufficient to promote transformation processes of carbides.

It may be surprising that radiation hardening effect in Haynes 230 is twice as lower as in Hastelloy X (Figure 12 and 3), while the increase of defects as a result of irradiation clearly exhibits an opposite tendency (Table 2). The mechanisms proposed in the previous paragraph shed new light on nanomechanical results' analysis. As highlighted before, GBS's presence significantly improves the Ni-based superalloys' creep strength. Nevertheless, recorded carbides and associated grain boundaries alteration suggest the early stages of GBS evolution. Thus, their beneficial influence on material mechanics cannot be noticed separately, although they may intensify the effects of other phenomena that can be noticed in presented results. The reasons underlying observed mechanical behavior's differences between those two materials are still related to structural features. It has been proved that the correlation between irradiation creep strength and yield strength of FCC materials is related to the kinetics of dislocation mobility [55,56]. In linear stress dependence of creep tests, dislocation climb is a major contributor to irradiation creep strain. High temperature and imbalance in point defect flux are factors promoting these mechanisms [57]. Furthermore, the relationship between yield strength and the material's hardness has also been confirmed [58]. Sun *et al.* [59] claimed that the level of stresses determines dominant creep mechanisms occurring during nanoindentation at room temperature. At lower stresses, activities on grain boundaries leading to lower strain rates are dominant. At higher stresses – dislocation activities prevail, and thus creep strain rate increases. In the presented experiment, the rate of loading at the regime of interest (investigation of the modified layer) may be regarded as low (approx. $\dot{P} \approx 0,05 mN/s$ – similarly to work [59]). Hence, deformation during the test is mediated by grain boundary activities. Since the surface fraction of grain boundaries is much lower in Haynes 230 than in Hastelloy X, activity on grain boundaries may be executed more easily in the first alloy. Obtaining a maximum effective load during tests leads to activating existing dislocations as well as nucleation of new ones [59]. Hence, in our case, larger grain allows for more effective nucleation and higher mobility of intragranular dislocation through the structure in Haynes 230. Fine grain and randomly dispersed small-size precipitates in Hastelloy X successfully limit the free distance for dislocation movement. Thus, the interaction between dislocations and existing structural obstacles hinders deformation resulting in higher hardness values. For this reason, despite the number of generated dislocations due to ion-irradiation being much lower in Hastelloy X than in Haynes 230 (Table 2), the first materials display a much more intensive hardening effect (Figure 14). Moreover, proposed mechanisms of partial decomposition on $M_6C$ carbides in the proximity of grain boundaries in Hastelloy X also may delay nucleation of new dislocation during nanoindentation experiments. Hence, a more intensive hardening effect recorded for this material is related to the structural features, processes occurring during ion irradiation, and finally, deformation parameters during indentation experiments. Still, it must be highlighted that higher hardness increment values may also be related to the improvement of creep strength by the evolution of GBS, as shown in works [13,49].

## 5. Conclusions

Radiation resistance is one of the most critical factors influencing the materials used in demanding applications, such as elements of IV gen. nuclear reactors. In this study, the two types of Ni-based superalloys designed for high-temperature long-term performance have been investigated. Commercially available Hastelloy X and Haynes 230 were characterized in the terms of structure and nanomechanical properties evolution. Both materials display FCC structure with numerous $M_{23}C_6$ and $M_6C$ carbides. However, grain size, precipitations morphology, and distribution significantly differ in both alloys. Presented work includes radiation damaging by Ar-ions up to 1.2 dpa and 12 dpa at 400ºC. Obtained results has revealed substantial radiation hardening of both materials; however response of Hastelloy X is more intensive. Moreover, carbides morphology change in this alloy has been recorded. It is believed that the early stages of grain boundary serrations (GBS) formation have been initiated due to radiation damage. It is known that GBS may be obtained in Ni-based alloys by appropriate heat treatment. This process is related to the progressing decomposition of $M_6C$ carbides, nucleation of $M_{23}C_6$, and γ' phase precipitates in grain boundaries. Observed effects suggest that ion irradiation has promoted processes preceding the evolution of GBS. The authors propose two reasons that may explain the observed effect:

(i) ion irradiation provides a sufficient energy repository to promote the decomposition of unstable at high-temperature $M_6C$ carbides and following nucleation of $M_{23}C_6$ precipitates through grain boundaries. Thus, these processes take place at a lower temperature regime than reported in the literature; and
(ii) increased mobility of defects (as a result of elevated temperature) and continuous generation of point defects during ion irradiation (that also beneficially influences diffusion processes) enhance processes leading to energy reduction more rapidly, such as the above-mentioned releasing atoms in decomposition processes and re-binding them into the new phases.

Moreover, similar effects are not recorded in the structure of Haynes 230. Fundamental differences in structural features, such as grain boundary volume, morphology, and distribution of precipitates, are believed to be reason for this observation. These elements determine the efficiency of basic mechanisms occurring during radiation damage, such as mutual recombination, diffusion to defect sinks, or generation of new dislocations. These mechanisms allow for sufficient energy reduction; thus, processes of carbide morphology change on grain boundaries are not promoted. These assumptions also explain differences in the estimated density of dislocation increase due to ion irradiation in both materials.

**Acknowledgements**

This work is one portion of the studies in the strategic Polish program of scientific research and development work "Social and economic development of Poland in the conditions of globalizing markets GOSPOSTRATEG" part of "Preparation of legal, organizational and technical instruments for the HTR implementation" financed by the National Centre for Research and Development (NCBiR) in Poland.

The authors acknowledge the support from the European Union Horizon 2020 research and innovation program under NOMATEN Teaming grant agreement no. 857470 and from the European Regional Development Fund via the Foundation for Polish Science International Research Agenda Plus program grant no. MAB PLUS/2018/8.